# Toward Co-adapting Machine Learning Job Shape and Cluster Topology

Shawn Shuoshuo Chen*    Daiyaan Arfeen*
Minlan Yu+    Peter Steenkiste*    Srinivasan Seshan*

*Carnegie Mellon University    +Harvard University

## ABSTRACT

Allocating resources to distributed machine learning jobs in multi-tenant torus-topology clusters must meet each job's specific placement and communication requirements, which are typically described using *shapes*. There is an inherent tension between minimizing network contention and maximizing cluster utilization when placing various-shaped jobs. While existing schedulers typically optimize for one objective at the expense of the other, we demonstrate that both can be achieved simultaneously.

Our proposed approach, *RFold*, adapts both job shapes and the underlying cluster topology at runtime. This is accomplished by combining two techniques: (1) identifying homomorphic job shapes that support the jobs communication needs, and (2) reconfiguring the optical circuit switch-enabled topology to support more diverse job shapes. Preliminary evaluation performed on a 4096-node torus cluster simulator indicates that RFold can improve absolute cluster utilization by 57% and reduce job completion time by up to 11× relative to existing methods.

## 1 INTRODUCTION

Machine learning (ML) clusters are pools of connected accelerators (also known as XPUs) typically shared by many concurrent jobs. Given that these clusters require substantial investment, there is a strong interest in their efficient operation. This efficiency depends on supporting as many jobs as possible on the cluster while minimizing the contention between jobs. Adding jobs to such clusters generally involves two sub-tasks: admission and placement. An admission policy determines whether to admit a job and how many XPUs to allocate, while the placement policy identifies the most suitable set of XPUs to use. Existing ML schedulers [13, 24, 28, 29, 31, 35, 38, 46, 48] primarily focus on admission, since placement is not a significant concern in Clos-topology GPU clusters. However, placement is an increasingly important problem as cluster designs move away from Clos topologies. In such settings, suboptimal placement often leads to severe resource underutilization and degraded job performance.

Recently, ML clusters based on torus topologies have been deployed in production [3, 19]. Unlike switched topologies, torus-based clusters are constructed with directly connected XPUs. This allows them to scale easily with minimal rewiring. Torus clusters also have lower construction costs compared to switched topologies while supporting the neighbor-centric communication patterns of typical ML workloads. On the other hand, only neighboring XPUs can communicate with non-blocking, full bandwidth. Data transmission between more distant XPUs must route through intermediate XPUs, leading to competition for link bandwidth and making a job's performance sensitive to its placement in the torus cluster.

Two main approaches have been proposed to handle job placement in torus clusters. The first employs a First-Fit algorithm (or its variants) [7] to search for a sufficient number of contiguous XPUs, and assigns a job to the first location discovered. The second approach attempts to find XPUs that are close to each other on a best-effort basis [22, 27], without guaranteeing contiguity. Both approaches suffer from significant drawbacks.

First-Fit often leads to resource fragmentation and low utilization in multi-tenant torus clusters. This issue arises because users submit jobs of varying sizes, which arrive and complete asynchronously. In addition, the communication patterns inherent to ML jobs also demand XPU allocations that conform to specific rectangular or cuboid shapes (§2). This shape constraint inevitably strands idle XPUs between allocations. Consequently, a new job might not be scheduled even if sufficient XPUs are available, simply because of non-contiguity. If no placement is found, the job must be queued until resources free up. The queueing delay can be pronounced since ML jobs typically last for long periods.

In contrast, best-effort placement, despite its higher utilization of cluster XPUs, may assign spatially scattered XPUs to a job. The resulting network contention is an especially prominent problem for ML workloads, since their constituent XPUs frequently exchange immense volumes of data. This contention prolongs job completion times.

We set out to create a more flexible cluster resource allocation, guided by two key insights. The first insight is that ML job communication patterns are regular and rearrangeable. A job's shape can be mapped to the torus topology in multiple contiguous ways without compromising performance. This insight inspires *folding*, a technique that exhaustively evaluates all feasible shape mappings and identifies the one that minimizes fragmentation. Our second insight reveals that adapting the topology to a job shape can significantly improve the chance of successful allocation. This can be accomplished using optical circuit switches (OCSes) [8, 40] to reconfigure the topology at runtime.

We explore the possibility of combining job shape alteration (folding) and topology alteration (reconfiguration) through the design of *RFold*, a new resource allocation scheme that aims to achieve high resource utilization while avoiding network contention. RFold attempts all legitimate shape variations for a given job—those that can limit network contention—by formulating shapes as graphs, and then invokes graph libraries to check for homomorphism. For each shape variation, it generates an allocation plan by virtually reconfiguring the underlying topology to best match the shape. Finally, RFold ranks these plans and commits to one that minimizes fragmentation.

To evaluate RFold, we developed a topology-aware, job-level simulator that simulates large-scale 3D torus clusters with reconfigurable OCS links. We also extend public ML job traces, such as the Microsoft Philly trace [17], based on our observations in a university-level cluster. We compare RFold to the First-Fit placement policy, as well as a folding-only policy and a reconfiguration-only policy. Our preliminary results suggest that RFold can reduce job completion time by up to 11× over the reconfiguration-only policy, increasing both the absolute cluster utilization by 57% and the success rate of job allocation by 10× over First-Fit. These findings show a strong potential for performance improvements through placement, topology design, and reconfigurability.

We provide background on: (1) the characteristics of ML jobs, (2) the difference between static torus and reconfigurable torus, and (3) the current status of ML job scheduling in §2. §3 describes how RFold works. Evaluation results on job performance and cluster utilization are presented in §4. We conclude and discuss some future directions in §5.

## 2 BACKGROUND

**3D parallelism.** ML training jobs often run on multiple XPUs to reduce running time. Three parallelization strategies[1] [32, 49] are commonly used to carry out distributed ML training: (1) data parallelism (DP), (2) tensor parallelism (TP), and (3) pipeline parallelism (PP). If the mini-batch of input data is too large to fit in the memory of a single XPU, DP shards it across multiple XPUs, with model replica per XPU. If the model itself is too large, TP and PP shard it across multiple XPUs. Many tools [18, 44, 49, 50] can determine the optimal combination of parallelization strategies for each ML job and how many XPUs to use for each dimension of parallelism. The resulting plan can be expressed as a job "shape" requirement. For example, a job with a 4×6×1 shape signifies that its model is sharded into six parts across six XPUs (six-way TP), and the model itself is replicated four times (four-way DP). This configuration results in two independent communication phases. One phase synchronizes the calculated model gradients across four DP-dimension XPUs corresponding to each model shard using six parallel ring-based AllReduce collective communication operations [33, 37]. The other phase synchronizes the six TP-dimension XPUs of each model replica with four parallel AllReduce operations. The DP and TP parallelism are orthogonal to each other, and their communication patterns are isolated to the participating XPUs. Similarly, a 18×1×1 shape indicates DP-only, and 4×4×4 denotes DP+TP+PP.

---

[1]Expert parallelism [23] and context parallelism [26] have also been proposed, but are not our focus in this paper.

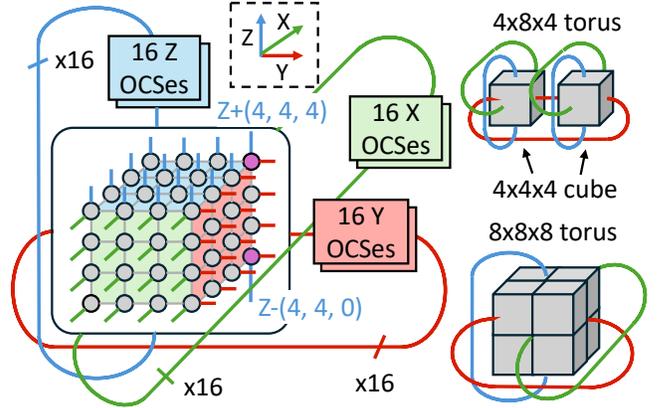

**Figure 1: An example of 3D torus cluster (Google TPU v4) built from 4×4×4 reconfigurable cubes.**

**Reconfigurable torus.** Since ML workloads heavily rely on ring-based communications, many existing and proposed system designs [2, 3, 10, 12, 15, 19–21, 25, 41] adopt a mesh or torus topology to match their communication pattern. The bandwidth demands of each parallelism differ, with TP being the most bandwidth-intensive and PP the least. Torus system designs provision for and ensure uniform worst-case bandwidth. Among these, earlier designs such as the Google TPU v2/v3 [20] and Amazon Trainium1 [2] clusters were based on 2D torus. As cluster scale expands, 3D torus becomes advantageous, as it offers lower network diameter and higher bisection bandwidth than 2D torus.

A notable constraint of statically wired 3D tori is the immutable nature of their topology, diameter, and dimensional sizes. In response, the Google TPU v4 cluster [19, 51] breaks a large 4096-XPU static torus into 64 hardwired cubes and introduces OCSes to dynamically reconfigure the links between the cubes. This not only improves availability but also proves to enable considerable flexibility for job placement.

As Figure 1 shows, each cube contains 4×4×4 XPUs, with 6 ports per XPU. The 16 ports on the opposite faces are connected to the same group of 16 OCSes. More specifically, two opposing ports at the same position are connected to the same OCS. For example, we label each XPU in a cube with an (X, Y, Z) coordinate, and the two Z-dimension ports of an XPU are named Z+ and Z-. Port Z+ of (4, 4, 4) and port Z- of (4, 4, 0) are connected to the same Z-dimension OCS. All 64 cubes are connected to these three groups of OCSes. OCSes allow the face ports of each XPU to either connect to the opposite face port of the same cube and form wrap-around links, or connect to the corresponding face port of another cube to form larger shapes, e.g., 4×8×4 and 8×8×8 tori.

**Job scheduling.** Ring AllReduce collective communication between separated XPUs requires traversing intermediate nodes (typically using dimension-order routing [30]), making network performance sensitive to intermediate XPU activity on shared links. While solutions like Cassini [36] and Crux [5] target network contention in GPU clusters, it remains an unsolved problem in torus clusters.

Another category of schedulers includes those designed for big data clusters [4, 14, 34, 39, 42, 43, 47]. They consider various resource types, e.g., CPU, GPU, RAM, disk, when making scheduling decisions. Among these, Borg [43], YARN [42], Mesos [14], Kubernetes [4] and SLURM [47] are used in production data centers. SLURM is also widely used in high-performance computing clusters and supports 3D torus topology. It uses a Hilbert curve [1] to map 3D nodes onto a 1D axis, so that XPUs with proximity can be found using line segment search algorithms. Unfortunately, none of them supports reconfigurable topologies.

Although job admission is beyond the scope of this paper, the research community has investigated a diverse set of admission policies. Optimus [35] and Gandiva [46] elastically adjust the number of XPUs allocated to each job to achieve the best JCT. Tiresias [13] relies on preemption to optimize JCT, whereas Themis [28] preempts jobs for improved fairness. Synergy [29], Gavel [31], Sia [38] and Lyra [24] all focus on scheduling heterogeneous compute resources but do not account for the interconnect. Our proposed placement policy could complement these existing admission policies.

## 3 RFOLD RESOURCE ALLOCATION

The question we seek to answer is: *how to achieve both contention-free job performance and efficient resource utilization in torus clusters? Can we leverage ML job characteristics and OCS reconfigurability to place jobs intelligently, and how?*

### 3.1 Motivation

To better understand the impact of suboptimal placement, we ran experiments on a TPU v2 machine in Google Cloud. We place a two-TPU ML job first on a row (ideal) of a 2×2 grid, and then on the diagonal line (non-ideal). The job's communication time in the diagonal configuration is 17% longer than that on a row. We then placed two identical ML jobs, each assigned 2 TPUs, on the two diagonal lines of the grid, forcing them to compete for a common link. This resulted in a 35% longer communication time compared to when a single job ran in the same diagonal configuration. When the one job's communication load was doubled (or tripled), the other measured job's communication time escalated by 95% (and 186%), respectively. These results show that the degree of performance degradation caused by suboptimal placement is unpredictable and can be very significant. Therefore, our goal is to avoid contention by enforcing the job shape so that allocated XPUs and links are always exclusive to the job.

In this section, we introduce *RFold*, a novel resource allocation scheme designed to address our problem. RFold combines two techniques, namely reconfiguration (§3.2) and folding (§3.3), to facilitate job placement with flexibility. Reconfiguration adapts the underlying cluster topology to match job shapes that would otherwise be challenging to place. Folding transforms original job shapes into equivalent forms while ensuring minimum network contention and resource fragmentation. We recognize that reconfigurable OCS links are more valuable than plain links, as they enable topology changes and defragmentation. Jobs should prioritize using plain links over OCS links. This insight leads to a core heuristic in RFold: the optimal placement consumes the fewest reconfigurable cubes and OCS links.

### 3.2 Reconfiguration

A static torus is commonly constructed with an equal number of XPUs in every dimension. For example, a 4096-XPU cluster would be built as 16×16×16. The static nature of this design limits the job shapes that can be supported. Consider a job that requires 4×4×32 XPUs. While the total number of XPUs required can be met, with the shape constraint, this job can never be placed because one of its dimensions exceeds the maximum dimension size of the torus (32>16).

Reconfigurable torus offers more flexibility in job placement, i.e., it can support more shapes by adapting the runtime topology to the job shape. Recall the TPU v4 cluster described in §2. It contains 64 4×4×4 hardwired cubes, and any two cubes can be connected via the OCSes to form a larger torus. Hence, to place the 4×4×32 job, we do not have to find 32 contiguous XPUs in one dimension. Instead, we only need eight 4×4×4 cubes to be reconfigured side-by-side.

When job shapes are not a multiple of four—for example, 4×4×34—it results in at least one partially used cube. To address this fragmentation, we could construct smaller cubes, like 1×1×1, to ensure full utilization. Yet, such a design would constrain the cluster's overall scale because if all torus links are OCS links, the OCSes would quickly run out of ports. We primarily focus on 4×4×4 cubes in our proposal since it is practical with current technology.

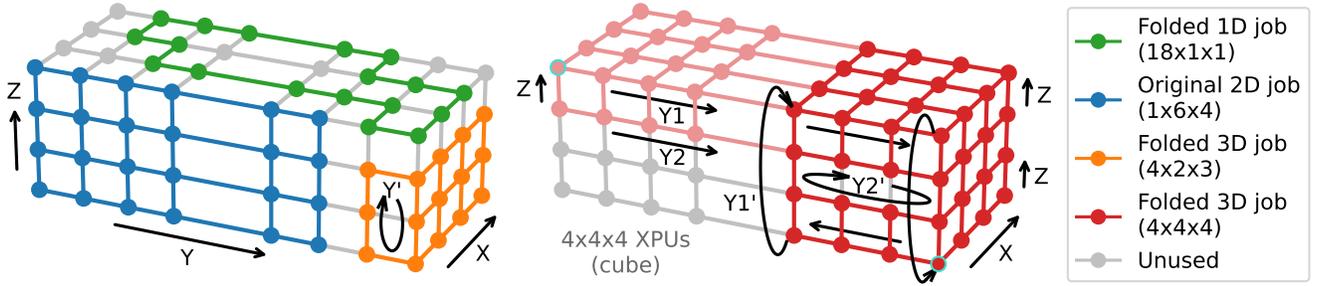

Figure 2: Demonstration of 1D, 2D and 3D job folding in a 3D torus. Each 4×4×4 cube is reconfigurable, with wrap-around links (not displayed) to itself on the X and Z axes, and to the other cube on the Y axis.

From a high level, the placement algorithm in a reconfigurable cluster works as follows: It generates multiple options to break down a large job shape into several smaller pieces and then searches for available resources in each cube that match the shape of a specific piece. After one or multiple feasible placements are found, we then use a simple heuristic to rank them and pick the one that requires the minimum number of cubes. Finally, we reconfigure the OCSes to connect these pieces without interfering with other jobs.

Although reconfiguration generally makes it easier to satisfy the job shape constraints, several inefficiencies remain. First, each cube is only reconfigurable on its faces. If a cube has available XPUs in the core but all face XPUs are used, its available XPUs cannot be connected to other cubes to host a larger job. Second, an XPU on the face can only connect to the corresponding XPU in another cube at the same position. Thus, even with enough available XPUs in two combined cubes, the two halves might not be allowed to connect if they are misaligned. Third, jobs in a reconfigurable torus only receive wrap-around links when their shapes are a multiple of the cube dimension size N. A reconfigurable torus makes N smaller (e.g., 16→4), allowing more jobs to utilize these wrap-around links. Nevertheless, in cases where the job shape is not a multiple of N, jobs still suffer from poor communication performance.

### 3.3 Folding

Folding changes the job shape to make the job easier to place, given the existing topology and resource availability. Thus, it complements reconfigurations and helps overcome reconfiguration inefficiencies. It is worth noting that rotation [9, 11] is a widely used technique in multi-dimensional bin packing. When placing a job within a 3D torus, rotation offers 3! = 6 distinct placement options while preserving the job shape. We assume rotation is a default behavior incorporated into all placement policies and is therefore not considered a specific aspect of folding.

When a job arrives, the scheduler leverages folding to find all the shape variants homomorphic to the original shape requested by a job. An original shape can be folded to another shape as long as the communication pattern can be faithfully mapped onto the new shape. Then the scheduler evaluates each of these shape variants by invoking reconfiguration. The shape variant consuming the least amount of cubes is chosen and committed. Since jobs in a 3D torus usually communicate in one, two, or all three dimensions, they are henceforth referred to as 1D/2D/3D (shaped) jobs. We illustrate how folding works for these jobs using the examples in Figure 2.

**1D folding.** 1D jobs have a shape of A×1×1, which indicates that the communication happens between A XPUs along a one-dimensional ring (e.g., data-parallel only). A naïve scheduler that treats all jobs in a 3D torus as 3D would allocate A XPUs along a straight line for these jobs. If A is not a multiple of cube size N, these jobs do not get wrap-around links, and cannot form a ring. In such unfortunate cases, the resource allocation and job shape mismatch would result in performance penalties. However, shape A×1×1 does not necessarily require all A XPUs to be placed on the same dimension. Rather, these A XPUs can be arbitrarily allocated as long as they form a cycle/ring. Therefore, the main task in 1D folding is to find a simple cycle of length A given the grid graph of currently available XPUs.

As an example, the green job in Figure 2 (left) is a 1D job of shape 18×1×1. There are only two available 4×4×4 cubes and they are reconfigured to form a larger 4×8×4 3D torus. No single dimension in this torus can provide 18 consecutive XPUs. With folding, we are able to find 18 scattered XPUs forming a cycle to meet the job shape requirement. Note that this example demonstrates how a 1D job is folded to 2D, but 1D jobs can also be folded to 3D in the same way. The only requirement is to find an A-XPU cycle.

**2D folding.** 2D jobs (of shape A×B×1) manifest a two-dimensional communication pattern, e.g., DP along one dimension and TP along the other dimension. For instance, the blue job in Figure 2 is a 2D job of shape 1×6×4, which requires the XPUs to communicate along the Y and Z dimensions. Though this 2D job can be placed as shown, it leads to two issues: (1) The 6 XPUs along the Y dimension are not a multiple of 4, so there is no wrap-around link. (2) There is resource fragmentation along the X dimension. The remaining 3 XPUs along the X dimension are unable to serve jobs of even shape sizes, which are the majority of jobs.

Folding can mitigate these two issues. Specifically, we can fold the original 2D job to a 3D job of shape 4×2×3 (orange job). This is feasible because shape 1×6×4 is graph-homomorphic to shape 4×2×3. Communication along the Z dimension in the original 2D job can be mapped to communication along the X dimension in the folded 3D job. Communication along the Y dimension in the original 2D job now becomes communication along the Y' direction (circular)—this is how folding forms a cycle for non-multiple-of-4 shape sizes even if there is no wrap-around link. This technique of forming a cycle is applicable to most jobs with even shape sizes. Note that a 2D job can also be folded to another 2D job. 2D-to-2D folding works similarly to 3D-to-3D folding. Next, we discuss how 3D-to-3D folding works.

**3D folding.** True 3D jobs have shape sizes greater than 1 in all three dimensions, and communicate in all dimensions. This characteristic makes them the least "foldable" compared to 1D and 2D jobs. Without loss of generality, it is easy to fold an M-dimensional job to an (M+1)-dimensional torus because the extra dimension can accommodate communication requirements from the lower dimensions. But folding an M-dimensional job in an M-dimensional torus is non-trivial because of the communication requirements.

Nonetheless, certain 3D jobs can be folded, such as the red 3D job in Figure 2 (right). The original 3D job has a shape of 4×8×2. To place it as is, we need two 4×4×4 cubes reconfigured to form a 4×8×4 torus. Half of the job will be placed in the left cube and the other half in the right cube. Through folding, it is possible to place the entire job in one single 4×4×4 cube. Essentially, the left half of the job (light red) is folded to the bottom half of the right cube, e.g., the front-top-left XPU maps to the front-bottom-right XPU.

The original shape 4×8×2 is homomorphic to the folded shape 4×4×4 with special communication mapping. More specifically, communication along the X dimension remains unchanged. Communication along the Z dimension also remains the same—the top two layers and bottom two layers of the cube communicate between themselves along the Z dimension. Hence, the Z-dimension links in the center between the middle two layers are not needed (greyed out). Communication along the Y dimension is mapped differently for the two layers of the original job shape. We label them Y1 and Y2. Y2 is mapped to Y2' (circular), similar to the 2D folding example. For communication along Y1, we cannot find a direct cycle like Y2'. So we take advantage of the wrap-around links and map Y1 to Y1'—the XPUs on the top and bottom layers first communicate along the Y dimension, then use the wrap-around links to communicate with each other without traversing the middle two layers.

On the other hand, a job of shape 4×8×3 *cannot* be folded to 4×4×6 in the same way because only the innermost two layers and the outermost two layers can form a cycle to communicate. In the case of 4×8×3, the middle layer cannot be mapped to any cycle in the 4×4×6 shape.

Overall, jobs can be ranked by their "foldability" in the following order: 1D>2D>3D. When placing multiple new jobs, we should prioritize 3D jobs and then 2D jobs because 1D jobs can fit into the available space with minimum effort.

## 4 EVALUATION

In this section, we demonstrate that RFold can improve job completion time by reducing network contention and making jobs more flexible to place. The evaluation is done using a custom job-level discrete event simulator. This simulator takes job traces and cluster topologies as inputs and reports statistics on each individual job and the cluster. We have constructed: (1) a reconfigurable 3D torus of 4096 XPUs comprising 64 4×4×4 cubes (referred to as the reconfigurable torus), (2) the same reconfigurable torus with 8 8×8×8 cubes, (3) the same reconfigurable torus with 512 2×2×2 cubes, and (4) a 16×16×16 4096-XPU static torus. Our experiments only investigate various placement policies and fix the admission policy as first-in, first-out (FIFO); that is, an unscheduled job will block all subsequent jobs. If a job cannot be scheduled because of its incompatible shape, the scheduler removes it from the system and proceeds to the next.

Since none of the publicly available ML job traces [6] were collected from a torus-based cluster, we take the job inter-arrival time and duration from the Microsoft Philly trace [17], and override the job size (number of required XPUs) by sampling from a truncated exponential distribution between 1 and 4096. We then generate the job shape using a custom probability distribution following this rule of thumb: small jobs (≤256 XPUs) are more likely to have a shape of 1D or 2D, while large jobs (>256) are usually 2D or 3D in shape. If a job size can be factorized into multiple shapes, we select one uniformly at random. This approach reflects our observation in an academic GPU-based cluster—most submitted jobs are small, and only perform DP and/or TP (1D or 2D). We leave it as future work to apply more recent traces, e.g., Alibaba PAI [45] and Acme [16].

| Policy | Avg JCR (%) |
|---|---|
| FirstFit ($16^3$) | 10.4 |
| Folding ($16^3$) | 44.11 |
| Reconfig ($8^3$) | 31.46 |
| RFold ($8^3$) | 73.35 |
| Reconfig ($4^3$) | 100 |
| RFold ($4^3$) | 100 |

Table 1: Average job completion rate over 100 runs. Higher is better.

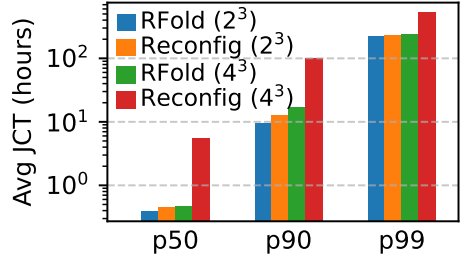

Figure 3: Job completion time at p50, p90 and p99, averaged across 100 runs.

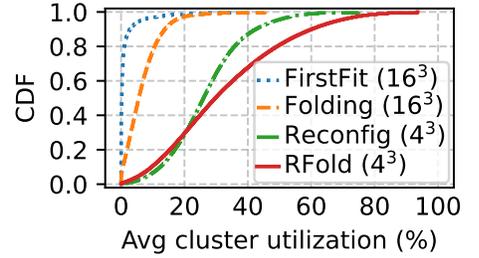

Figure 4: Cluster utilization. Each percentile is averaged across 100 runs.

We focus on three key metrics: job completion rate (JCR), job completion time (JCT), and cluster utilization. Table 1 lists the JCR—the ratio of successfully scheduled jobs to the total jobs—of different placement policies. For each policy, we repeat the simulation 100 times with 100 generated traces and report the average.

FirstFit runs in the static torus. It can only schedule 10.4% of the total jobs. This is because many jobs have one or more dimensions in their shapes that exceed the corresponding dimension of the static torus, preventing them from ever being scheduled. In other words, FirstFit cannot support jobs with a diverse set of shapes. The folding-only policy (Folding) also runs in the static torus. Its JCR increases to 44%, a significant improvement over FirstFit. The remaining 56% of jobs cannot be scheduled due to the same shape incompatibility. Folding's higher JCR is primarily due to 1D and 2D folding, which allows the incompatible shapes to be folded to fit. In contrast, 3D folding provides no benefit in a static torus, as it usually lacks the wrap-around links at 4×4×4 granularity.

The reconfiguration-only policy (Reconfig) breaks jobs into multiple pieces, allowing them to be placed into separate cubes while maintaining the appearance of their original shapes. This provides more freedom of placement compared to the static torus. As a result, a reconfigurable torus with 8×8×8 cubes achieves 31% JCR. If shrinking the cube size to 4×4×4 (or smaller), JCR further improves to 100%. RFold attains a 73% (or 100%) JCR with 8×8×8 (4×4×4) cubes.

Figure 3 reports the median (p50) and tail (p90, p99) JCT results averaged across 100 measurements. Since JCT is only meaningful when 100% of the jobs are succesfully scheduled, we only look at Reconfig and RFold with cube size no greater than 4×4×4. As shown, with 4×4×4 cubes, RFold outperforms Reconfig with 11×, 6× and 2× shorter JCT at p50, p90 and p99, respectively. A smaller cube size (2×2×2) allows for finer-grained topology reconfiguration. Hence, Reconfig performs more efficiently with these smaller cubes. Nevertheless, RFold still outperforms Reconfig, achieving up to 1.3× shorter JCT.

The cluster utilization when running different placement policies is presented in Figure 4. We sample cluster utilization in each run as a time series, and plot it as a CDF. Both FirstFit and Folding struggle to keep more than 40% of the XPUs busy. We attribute their low utilization to the insufficient number of scheduled jobs. Folding has a higher JCR than FirstFit, thereby improving the cluster utilization by up to 10% over FirstFit. Compared to the non-reconfigurable counter-parts, Reconfig and RFold both achieve significantly higher utilization owing to their reconfigurability. RFold further improves cluster utilization by an additional 20% over Reconfig because of job folding.

## 5 CONCLUSION

Job placement and resource allocation in torus-based ML clusters is challenging. ML jobs require specific shapes to accommodate their parallelization strategies. Suboptimal job placement leads to resource fragmentation and network contention. We attempt to tackle this problem with a novel technique called RFold. RFold leverages OCSes to reconfigure the cluster topology at runtime in order to adapt to the job shapes. Meanwhile, it explores different job shape variants to minimize contention. Our early results demonstrate RFold's promising performance in JCR, JCT, and cluster utilization. Despite RFold's potential for efficiency improvements, several open questions call for further exploration.

**Revisiting best-effort placement.** While this paper emphasizes contiguous job placement, the best-effort approach can still be considered a viable option. More specifically, starting a job immediately with a non-contiguous placement is acceptable as long as the slowdown from network contention is less than the queueing delay incurred by waiting for the next available contiguous placement.

**Reconfigurability.** We have discussed one practical reconfigurable torus topology based on 4×4×4 cubes (§3). Nevertheless, designs incorporating cubes of alternative sizes might prove beneficial given different optimization objectives. Larger cubes support more scalable clusters, especially considering constraints on OCS port counts. On the other hand, smaller cubes provide finer-grained reconfigurability, which could further improve job performance.

**Beyond 3D.** The focus of this paper is 3D torus topology and job shapes of 3 dimensions or less. While these are reasonable design choices considering current technology and practice, they are by no means fundamental. Future work should consider applying reconfiguration and folding to other topologies and jobs with higher dimensional shapes. *This work does not raise any ethical concerns.*